# The Accuracy of Mass Determination in Galaxy Clusters by X–ray Observations

S. Schindler[1,2,*]

[1] Max-Planck-Institut für extraterrestrische Physik, Giessenbachstraße, D-85478 Garching, Germany
[2] Max-Planck-Institut für Astrophysik, Karl-Schwarzschild-Straße 1, D-85478 Garching, Germany

**Abstract.** We test the reliability of mass determination in clusters of galaxies by X-ray observations. The true mass in cluster models is compared to the mass derived by the X-ray emission and X-ray temperature of a model assuming hydrostatic equilibrium and spherical symmetry. In general we find good agreement between the X-ray mass and the true mass without any systematic over- or underestimation. The average deviation within the central 2Mpc has a r.m.s. of 15%.

## 1. Introduction

Clusters of galaxies are the most massive, clearly defined objects in the universe. A precise knowledge of their mass is important for several astrophysical and cosmological questions. From a comparison of a virial mass estimate of the Coma cluster with an estimate of the mass in luminous galaxies Zwicky (1933) derived the first evidence of the existence of a dark matter component on large scales. Today the derivation of detailed gravitational mass profiles – e.g. from X-ray observations – allow the conclusion that this dark matter component must be smoothly distributed throughout the cluster and cannot, for example, be concentrated in one compact central object or completely associated with the massive galactic halos (David et al. 1994; Böhringer 1994).

Another important parameter that relies on the precise determination of the gravitational mass is the gas mass to total mass ratio. The fact that clusters of galaxies are the deepest, large-scale gravitational potentials in the universe in which dark matter, galaxies, and gas are accumulated indiscriminately has been used to infer the baryon to dark matter ratio in clusters (White & Frenk 1991; Böhringer et al. 1993). White et al. (1993b) have pointed out that the baryon to dark matter ratio (of $\approx$ 10-30%) in clusters if extrapolated to the universe as a whole is in conflict with the results of primordial nucleosynthesis models for cosmological models with $\Omega_0$=1.

[*] e-mail: sas@mpa-garching.mpg.de

Clusters of galaxies are also important probes for the large-scale structure in the universe. In particular, the cluster mass function can be used to constrain the power spectrum of the primordial density fluctuations at comoving scale of 5 – 15 Mpc (e.g. Henry & Arnaud 1991; Böhringer & Schindler 1992; White et al. 1993a; Bahcall & Cen 1993).

All these applications rely on a precise determination of the cluster mass. The first approach was based on the spatial distribution and velocity dispersion of the galaxies (The & White 1986; Merritt 1987). But it was found that the uncertainty of the degree of anisotropy in the galaxy orbits and the poor number statistics of galaxies with observed redshifts does not allow for a tight constraint of the cluster mass profile (The & White 1986; Merritt 1987).

X-ray observations of the intracluster plasma provide a more reliable approach. Since the sound crossing time in the intra-cluster medium (ICM) within an Abell radius of 3 Mpc is only 1/6 - 1/3 of the Hubble time it is generally assumed that the ICM is in approximate hydrostatic equilibrium. (Throughout the paper we use a Hubble constant of $H_0 = 50 \mathrm{km\ s^{-1}Mpc^{-1}}$.) With the knowledge of the ICM density and temperature distribution from X-ray observations one can determine the gravitational mass profiles by means of the hydrostatic equation. Further simplifications – e.g. assumption of spherical symmetry – also enter the mass determination in practice.

Recently, it has also become possible to determine cluster masses from the effect of gravitational lensing. Either the lens modeling of giant gravitational arcs or the weak lensing effect is used to obtain gravitational potential models for clusters (e.g. Kneib et al. 1993; Kaiser & Squires 1993). In some recent results it appears as if the mass determination from gravitational lensing would lead to considerably higher masses than the conventional approaches based on galaxy velocity dispersions or X-ray observations (Miralda–Escudé & Babul 1994; Fahlman et al.1994). This rises the question how reliable the different models really are.

In this paper we therefore investigate the robustness of mass determination based on X-ray observations. We compare the true input masses of simulated cluster models with the mass obtained from simulated X-ray observations of these model clusters with subsequent mass determination. In particular, it is analysed in detail how the assumption of hydrostatic equilibrium and geometric simplifications (e.g. neglect of substructure) influence the final result.

In Sect. 2 we show a comparison of hydrodynamic and hydrostatic cluster models. A short description of the method for the mass determination is given in Sect. 3. The results are presented in Sect. 4 and discussed in Sect. 5.

## 2. Comparison of hydrodynamic and hydrostatic models

Hydrostatic cluster models are used for a wide range of applications. Therefore we show at first the differences between hydrodynamic and hydrostatic models. For this comparison we choose a configuration in which two subclusters have merged recently and a shock wave is emerging. At this stage the strongest deviations from hydrostatic equilibrium are expected. Therefore this configuration represents an upper limit for the errors that are introduced with the assumption of hydrostatic equilibrium.

We use the hydrodynamic models described in Schindler & Böhringer (1993) and Schindler & Müller (1993) and an additional model which includes more substructure. The models are calculated with a combined N-body and hydrodynamic method. While for the N-body method a direct integration scheme (Aarseth 1972) is used, the hydrodynamics is calculated with the Piecewise Parabolic Method (Fryxell et al. 1989) – a grid code which is specially designed to treat shock waves correctly.

To construct a hydrostatic model corresponding to each hydrodynamic model we use the potential of the hydrodynamic model and fill gas into this potential in hydrostatic equilibrium with a polytropic stratification with the same polytropic index that is used for the initial conditions of the hydrodynamic model. The polytropic indices in the different models vary between 1.01 and 1.3. The total gas mass is the same in two corresponding models.

The X-ray appearance of a model cluster (Fig. 1a) is calculated by integrating the emission along the line-of-sight and convolving it with the effective detector area of the ROSAT/PSPC. For the X-ray temperature (Fig. 1b) an emission weighted temperature average is calculated along the line-of-sight.

The model cluster in Fig. 1 has recently undergone a collision with a subcluster. The collision axis is oriented from upper left to lower right. During the collision gas is ejected from the cluster perpendicular to the collision axis, while along the collision axis gas is still flowing towards the centre of the cluster. This asymmetric gas ejection is reflected in the shape of the resulting shock front. This lens-shaped shock front is visible in ROSAT count rate images (Fig. 1a) as well as in the X-ray temperature maps (Fig. 1b).

The corresponding hydrostatic images are shown in Figs. 1c and 1d. The X-ray appearance is calculated in the same way as it is done for the hydrodynamic model. The hydrostatic images look very round and smooth compared to the hydrodynamic images because all the shocks are missing.

To stress the differences more clearly we show a ratio image of Figs. 1a and 1c (Fig. 1e) and a ratio image of Figs. 1b and 1d (Fig. 1f). The lens-shaped shock front is clearly visible in the count rate ratio image (Fig. 1e) and even more pronounced in the temperature ratio image (Fig. 1f). The largest deviations in the count rate are in the centre of the cluster and at the positions of the strongest shocks. Here the X-ray emission is overestimated by the hydrostatic model. The temperature ratio image (Fig. 1f) shows that the hydrostatic model gives too small temperatures behind and slightly too large temperatures in front of the shock fronts.

## 3. Mass determination method

For the mass determination from X-ray observations two assumptions are made: hydrostatic equilibrium and spherical symmetry. These two assumptions and the equation of state for an ideal gas yields the formula for the integrated mass

$$M(r) = \frac{kr}{\mu m_p G} T \left( \frac{d \ln \rho}{d \ln r} + \frac{d \ln T}{d \ln r} \right), \qquad (1)$$

where $\rho$ and $T$ are the density and the temperature of the intracluster gas. $r$, $k$, $\mu$, $m_p$, and $G$ are the radius, the Boltzmann constant, the molecular weight, the proton mass, and the gravitational constant.

For determining the density and the temperature profile we use three different methods with increasing degree of simplicity: In the first method we use the full information of the three-dimensional models. The second method is the one that is usually applied to X-ray observations: the two-dimensional observed data are deprojected with the assumption of spherical symmetry to infer the three-dimensional distribution. The third method is the simplest way to determine masses from X-ray observations: the two-dimensional quantities – without any deprojection – are used to determine the profiles.

In the first method an average density and an average temperature is calculated for radial shells using three-dimensional data. From these averaged quantities the radial density and temperature gradients are calculated. The mass derived from the mass formula (Eq. (1)) with these values inserted is compared with the true mass of the model. This analysis tests how good the assumptions of hydrostatic equilibrium and spherical symmetry are.

Furthermore, we analysed the models in the way that observed data are treated. Unfortunately, the three-dimensional quantities are not available from observations. Therefore the observed two-dimensional X-ray data have to be deprojected. There are two approaches for this deprojection which both assume spherical symmetry: a numerical deprojection as it is used e.g. in Nulsen & Böhringer (1995) and a deprojection by fitting the two-dimensional data with a $\beta$-model (Cavaliere &

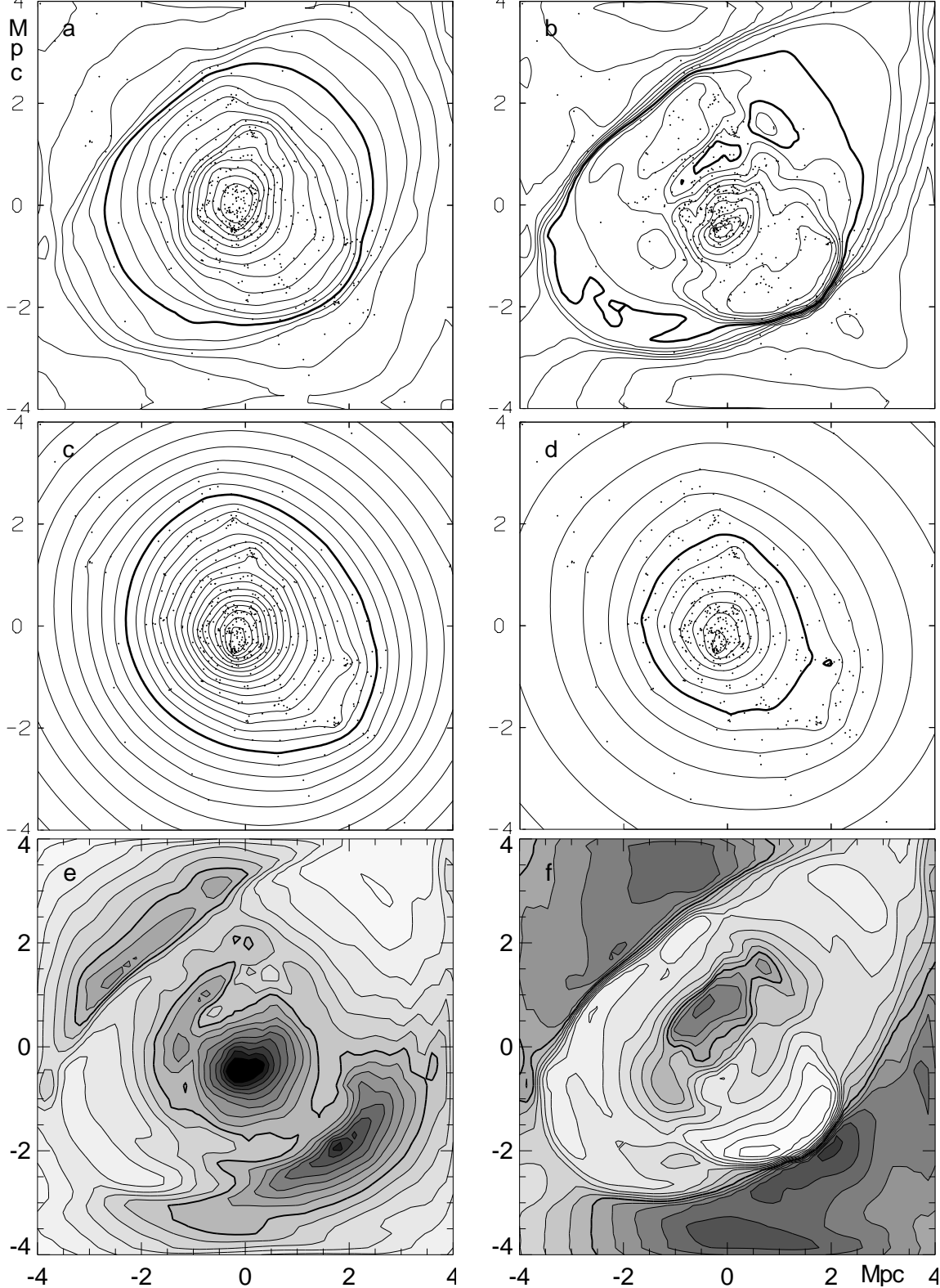

**Fig. 1.** X-ray images of a simulated cluster. Contours of constant count rate of a hydrodynamic (a) and a hydrostatic (c) model cluster with a logarithmic spacing with five contours per decade. The bold contour line corresponds to a count rate of $10^{-8} \mathrm{s}^{-1} \mathrm{kpc}^{-2}$. The second column shows X-ray temperature maps of the same hydrodynamic (b) and hydrostatic (d) cluster. The contours are logarithmically spaced with $\Delta \log T = 0.05$. The bold contour line corresponds to a temperature of $10^8$K. e) ratio of (a) and (c). The contours are logarithmically spaced with 10 contours per decade. In the dark regions the count rate of the hydrostatic model is too high. f) ratio of (b) and (d). The contours are logarithmically spaced with 25 contours per decade. In the dark regions the temperature of the hydrostatic model is too high. The bold contour line in (e) and (f) corresponds to a ratio of 1.

ness, the core radius, and $\beta$ (the ratio of the kinetic energy of the galaxies to the thermal energy of the gas) can be derived as fit parameters. With these parameters the three-dimensional density distribution is determined.

In a first step we project the three-dimensional models to two-dimensional images that ROSAT would have obtained by observing the model clusters. These images are then fitted with a $\beta$-model to obtain the density profile. The temperature profile is more difficult to derive from observations because the temperatures from spectral fits have relatively large error bars. A deprojection analysis for the temperature results in even larger errors and is therefore not necessary for many clusters (see e.g. Neumann & Böhringer 1995). Therefore, for most observations the two-dimensional temperature profile is used. Nevertheless, we use in this paper the three-dimensional temperature distribution to avoid the systematic errors that would be introduced by this simplification. These errors are discussed later.

For comparison we derive masses with a third, very simple method: We calculate along the line-of-sight projected, emission-weighted densities and temperatures. (The observational equivalent to the emission-weighted density is the square root of the surface brightness. A normalisation is not necessary because only relative gradients are needed.) These two-dimensional data are radially binned and subsequently treated as if it were the three-dimensional density and temperature profiles.

## 4. Results

We divide the investigation into two parts. In the first part we analyse so-called "normal" clusters – the clusters to which typically the mass determination is applied. The second part presents some extreme situations – like clusters with shock waves or substructure – which can affect the mass determination.

### 4.1. Normal clusters

The models that we call "normal clusters" are models where no obvious subcluster or group of galaxies is falling into the main cluster. These model clusters are almost in virial equilibrium, although they are not exactly spherically symmetric. There is some motion of collisionless matter and of intracluster gas in these clusters which causes small bumps in the X-ray images. They are very similar to the images of observed clusters. We applied the analysis to 21 model configurations from 6 different simulations with different initial conditions and different resolution.

In the three-dimensional analysis the agreement between the true mass and the mass derived by the hydrostatic method is generally quite good. A typical mass profile is shown in Fig. 2. There are only some small deviations at varying radii which are typically caused by varying gradients connected to the above mentioned bumps.

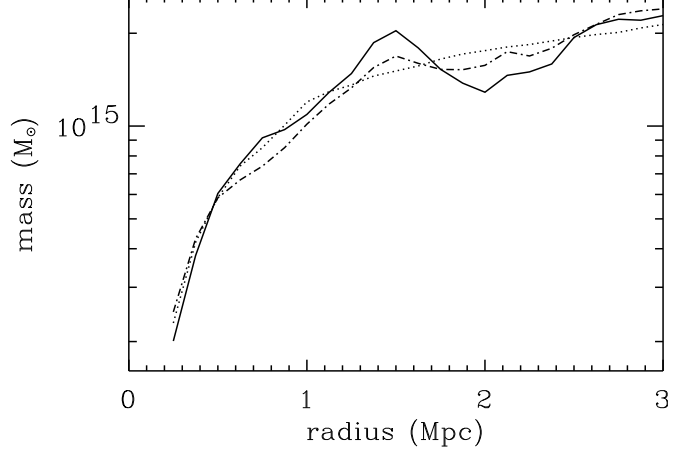

**Fig. 2.** Typical mass profile of a cluster that is roughly in virial equilibrium. The true integrated mass (dotted line) and the mass derived by the hydrostatic assumption from three-dimensional data (full line) and from X-ray images (dash-dotted line) are plotted versus radius. In general, there is good agreement between the two curves.

As the hydrostatic mass is not derived by a summation it is possible that – although an integrated mass is plotted – the mass is decreasing occasionally. We cannot compare the masses within the core region of the clusters due to the limited resolution of the hydrodynamic models.

Averaging over all deviations of hydrostatic mass from true mass in all model configurations we find on average an underestimation of the mass of only 2%. The deviations show a small tendency to increase with radius. That means there is no intrinsic error in the method that would cause a systematic over- or underestimation of the mass. Furthermore, no dependences on the initial conditions or the resolution of the simulated models can be found.

Calculating the r.m.s. of the deviations we find the same trend of increasing deviation with increasing radius. Within a radius of 2Mpc we find on average a standard deviation of 15%, within 3Mpc a standard deviation of 18%. Also for the standard deviations we do not find any dependences on the initial conditions or the resolution of the simulated models.

The second method – the analysis including a deprojection with a $\beta$-model – gives basically the same results. An example is shown in Fig. 2. In some parts where the first method shows bumps in the mass profile the second method gives an even better mass estimate. The reason why these bumps are smoothed out in the $\beta$-method is that the fit can not follow the bumps in the density profile. All the remaining bumps are caused by the temperature profile which typically plays a minor role.

Averaging over all model configurations and all radii up to 3Mpc gives a mean deviation from true mass of less than 1%. This method shows a trend to underestimate the mass by $\approx 10\%$ for radii smaller than 1.5Mpc and to overestimate the mass, also by about 10%, for radii larger than 1.5Mpc. Obviously, the $\beta$-model gives slightly too small density gradients for small radii and vice versa.

The r.m.s. of the deviations are for most models between 10 and 15%. Only some special models which where started with an almost isothermal distribution ($\gamma$=1.01) show larger deviations of 20-35%. In these models temporary cool regions ($\approx 10^7$K) develop which are caused by rarefaction waves propagating behind shock fronts. These cool regions cause bumps in the temperature profile, which were almost compensated by the density distribution in the three-dimensional analysis, but cannot be compensated by the simple fitted profile.

In contrast to the three-dimensional temperature distribution that we use in this analysis the projected two-dimensional temperature distribution is often used in the analysis of observational data. The effect of this projection is that the temperature gradients are smaller which causes a mass underestimation for a decreasing temperature profile and an overestimation for an increasing temperature profile. But as the dominant gradient in Eq. (1) is found to be always the density gradient the error of this projection is negligible. The density gradient is typically a factor of 2-30 larger than the temperature gradient.

The third, very crude method that is using only the two-dimensional non-deprojected data yields a systematic underestimation of the mass. The reason for the underestimation is that the density gradients are shallower because of the projection procedure and that gives – according to Eq. (1) – a smaller mass. The mean deviation within a radius of 3Mpc is between 10 and 20%. When only smaller radii are taken into account the projection effects are larger and therefore the mass underestimation is worse: for radii up to 0.5Mpc the deviation is between 20 and 30%.

### 4.2. Extreme cases

We investigate also some extreme cases – extreme in the sense that there is a strong deviation from hydrostatic equilibrium or from spherical symmetry.

#### 4.2.1. Shock waves

A strong deviation from hydrostatic equilibrium occurs e.g. when a strong shock wave emerges after a subcluster collision. In these cases we find an overestimation of the mass at the radius of the shock wave. The overestimation is caused by the large gradients at the shocks and large gradients give – according to Eq. (1) – large masses. For the strongest shock waves in our models we find locally overestimations of more than 100%. Figure 3a shows the mass profile of the model configuration shown in Figs. 1a and 1b which includes a very pronounced shock (note that the displayed mass range is much larger than in Fig. 2). The maximum deviation from the $\beta$-model analysis is even smaller than the deviation in the three-dimensional analysis because the fitting procedure has a smoothing effect on the profile.

#### 4.2.2. Subclustering

Deviations from spherical symmetry occur when substructure is present in a cluster. To investigate the influence of substructure on the mass determination we use model configurations with subclusters and small groups of galaxies. Only hydrostatic models are investigated to test the influence of deviation from spherical symmetry alone without any hydrodynamic effects.

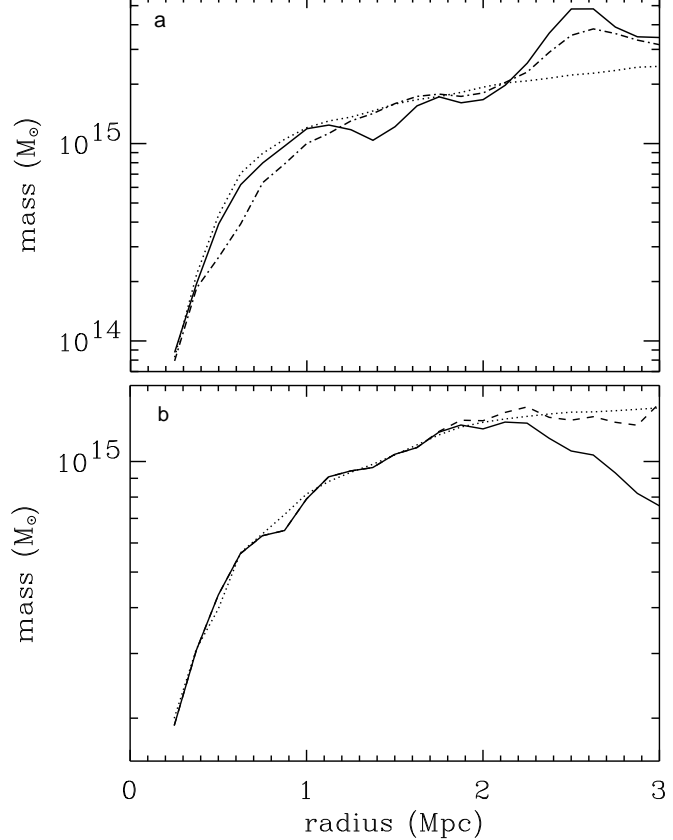

**Fig. 3.** Mass profiles of a model with strong shock waves (a) and a hydrostatic model with pronounced substructure (b). The true integrated mass (dashed line) and the mass derived by the hydrostatic assumption from three-dimensional data (full line) are plotted versus radius. The dash-dotted line in (a) is the mass profile derived by deprojection with a $\beta$-model. The dashed line in (b) shows the hydrostatic mass profile after removing the disturbed parts of the cluster.

The effect of subclustering is typically an underestimation of the mass around the radius of the substructure (see Fig. 3b). The density gradient is shallower when subclusters contribute to some outer shells and a shallower gradient gives – according to Eq. (1) – a smaller mass. In cases of very pronounced substructure we find even negative masses by applying Eq. (1) which is of course not physical.

The problem of mass determination in clusters with substructure is well known by X-ray astronomers. The solution is typically to clip the critical parts of the cluster and apply the method only on the undisturbed part (see e.g. Henry et al. 1993). Therefore we applied the same clipping to the model clusters and found good agreement of the mass for the rest cluster (see Fig. 3b). This example shows that many disturbed clusters which fall at first into the category of extreme case can

## 5. Discussion and conclusions

We compare the true cluster masses with the mass derived from simulated X-ray observations under the assumption of hydrostatic equilibrium and spherical symmetry. For roughly virialised clusters we find relatively small mass deviations. Using the three-dimensional data we get on average a r.m.s. of the deviations of 15% within a radius of 2Mpc. Starting from two-dimensional data and deprojecting the density distribution with a $\beta$-model fit yields a similar result. With these two analysis methods no general under- or overestimations of the mass were found. We tried a very crude third method which uses only two-dimensional, non-deprojected data. Even with this crude method which is much simpler than the analyses that observers usually apply we find on average a systematic underestimation not larger than 10-20% within the central 2 Mpc.

These errors are much smaller than today's overall errors introduced by observational uncertainties, e.g. typical errors in the temperature of the intracluster gas are 50% (e.g. Neumann & Böhringer 1995). That means that the errors introduced by the assumptions of hydrostatic equilibrium and spherical symmetry are in most clusters negligible.

Our results are in good agreement with investigations by Evrard (1994) who also compared X-ray mass estimates with the true masses of model clusters. By applying larger simplifications – he assumed an isothermal cluster – he found an accuracy of the X-ray mass of 50%.

We find the temperature gradient to be typically much smaller than the density gradient. This is advantageous for the analysis because the temperature gradient is difficult to determine from X–ray observations as already the temperatures have large errors. Because the mass is estimated from a sum of the dominant density gradient and the temperature gradients (Eq. (1)) a badly determined temperature gradient plays only a minor role.

We conclude that the mass determination from X-ray data with the assumption of hydrostatic equilibrium and spherical symmetry is in general very reliable. There may be some extreme cases where a deviation of a factor of two is possible: Strong shock waves in a cluster could lead to an overestimation of the mass. Substructure can cause a substantial underestimation of the mass. These over- or underestimations do not occur for the whole cluster, but only at specific radii. In some clusters such deviations have been found by comparing the X-ray mass with the mass derived from the gravitational lens effect with the X-ray mass typically being smaller, e.g. in A2218 (Miralda–Escudé & Babul 1994). The mass estimate from strong lensing – such as in the case of A2218 – is determined only at a specific radius, that is the distance of the arc from the cluster centre. A2218 is not a relaxed cluster but it is obviously disturbed (Kneib et al. 1995). Therefore, it is possible, that the X-ray mass estimate is too low because of some substructure. In general, substructure increases the probability for background galaxies to be lensed by clusters, which could explain the tendency that in lensing clusters the X-ray mass estimate is on average lower than the lensing mass.

Despite of some deviations that can show up in single clusters the method is very accurate for a larger sample of clusters where the deviations cancel out.

*Acknowledgements*. I am grateful to Hans Böhringer for many suggestions. Doris Neumann is thanked for useful discussions and Maximilian Ruffert for visualisation assistance. I thank B.A. Fryxell and E. Müller for kind permission to use their hydrodynamic programme and S.J. Aarseth for his N-body code. I gratefully acknowledge the hospitality of the Astronomical Institute of the University of Basel and thank the Verbundforschung for financial support.